\documentclass[sigconf]{acmart}
\usepackage{booktabs}  
\usepackage{arydshln}
\usepackage[most]{tcolorbox}

\newif{\ifhidecomments}
 \hidecommentsfalse 
\ifhidecomments 
    \newcommand{\janice}[1]{}  
    \newcommand{\wenhan}[1]{} 
    \newcommand{\yifan}[1]{} 
\else 
    \newcommand{\janice}[1]{\textbf{\sffamily{\textcolor{purple}{[Janice: #1 ]}}}} 
    \newcommand{\wenhan}[1]{\textbf{\sffamily{\textcolor{olive}{[Wenhan: #1 ]}}}}  
    \newcommand{\yifan}[1]{\textbf{\sffamily{\textcolor{cyan}{[Yifan: #1 ]}}}}  
\fi

\AtBeginDocument{%
  }

\copyrightyear{2025}
\acmYear{2025}
\setcopyright{cc}
\setcctype{by}
\acmConference[L@S '25]{Proceedings of the Twelfth ACM Conference on
Learning @ Scale}{July 21--23, 2025}{Palermo, Italy}
\acmBooktitle{Proceedings of the Twelfth ACM Conference on Learning @ Scale(L@S ’25), July 21--23, 2025, Palermo, Italy}
\acmDOI{10.1145/3698205.3729544}
\acmISBN{979-8-4007-1291-3/2025/07}

\newcommand{\takeaway}[1]{
\noindent\rule{\linewidth}{0.1pt}
\par\nobreak\noindent\textbf{Takeaways:}
#1
\vspace{-2mm}
\par\nobreak\noindent
\rule{\linewidth}{0.1pt}
}

\newtcolorbox{compactquote}{
    colback=white,           
    colframe=white,           
    boxrule=0pt,              
    borderline west={2mm}{-2mm}{black}, 
    left=3pt, right=0pt,       
    top=0pt, bottom=0pt, 
    boxsep=0pt, 
    before skip=5pt, after skip=5pt
}
\newcommand{\quoteblock}[2]{%
    \begin{compactquote}
        \textit{#1} (#2)
    \end{compactquote}
}

\newcommand{\pp}{\textbf{PP}\protect\includegraphics[height=2mm]{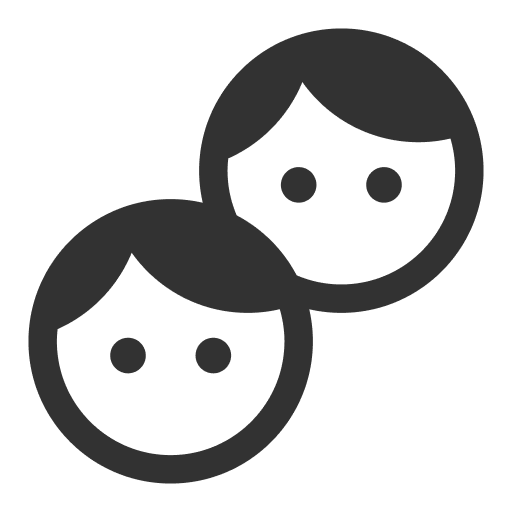}}
\newcommand{\ip}{\textbf{IP}\protect\includegraphics[height=2mm]{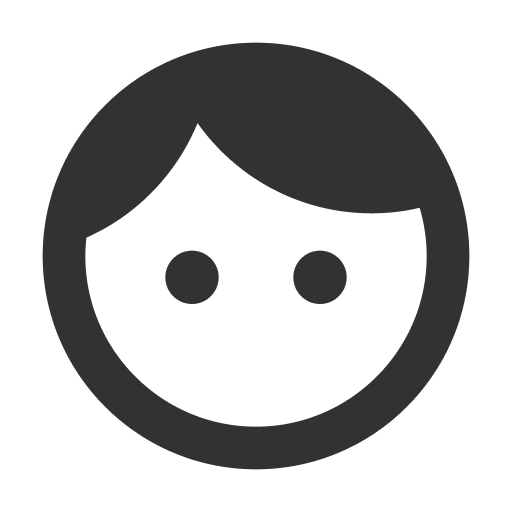}}
\newcommand{\pai}{\textbf{PAI}\protect\includegraphics[height=2mm]{figures/two.png}\protect\includegraphics[height=2mm]{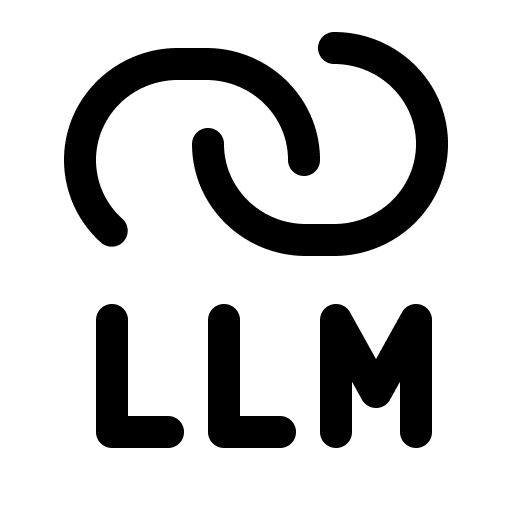}}
\newcommand{\sai}{\textbf{SAI}\protect\includegraphics[height=2mm]{figures/one.png}\protect\includegraphics[height=2mm]{figures/llm.png}}

\begin{document}

\title[Will Your Next Pair Programming Partner Be Human?]{Will Your Next Pair Programming Partner Be Human? An Empirical Evaluation of Generative AI as a Collaborative Teammate in a Semester-Long Classroom Setting} 



\author{Wenhan Lyu}  
\affiliation{
    \institution{William \& Mary}
    \city{Williamsburg}
    \state{VA}
    \country{USA}} 
\email{wlyu@wm.edu}
\orcid{0009-0004-9129-8689} 

\author{Yimeng Wang}  
\affiliation{
    \institution{William \& Mary}
    \city{Williamsburg}
    \state{VA}
    \country{USA}} 
\email{ywang139@wm.edu}
\orcid{0009-0005-0699-4581} 

\author{Yifan Sun}  
\affiliation{
    \institution{William \& Mary}
    \city{Williamsburg}
    \state{VA}
    \country{USA}}
\email{ysun25@wm.edu}
\orcid{0000-0003-3532-6521}

\author{Yixuan Zhang}  
 \affiliation{
    \institution{William \& Mary}
    \city{Williamsburg}
    \state{VA}
    \country{USA}} 
\email{yzhang104@wm.edu}
\orcid{0000-0002-7412-4669}
 
\begin{abstract}
Generative AI (GenAI), especially Large Language Models (LLMs), is rapidly reshaping both programming workflows and computer science education. Many programmers now incorporate GenAI tools into their workflows, including for collaborative coding tasks such as pair programming. While prior research has demonstrated the benefits of traditional pair programming and begun to explore GenAI-assisted coding, the role of LLM-based tools as \textit{collaborators} in pair programming remains underexamined. 
In this work, we conducted a mixed-methods study with 39 undergraduate students to examine how GenAI influences collaboration, learning, and performance in pair programming. 
Specifically, students completed six in-class assignments under three conditions: \textit{Traditional \textbf{P}air \textbf{P}rogramming} (\pp), \textit{\textbf{P}air Programming with Gen\textbf{AI}} (\pai),  and \textit{\textbf{S}olo Programming with Gen\textbf{AI}} (\sai). They used both LLM-based inline completion tools (e.g., GitHub Copilot) and LLM-based conversational tools (e.g., ChatGPT). Our results show that students in the \pai{} condition achieved the highest assignment scores, whereas those in the \sai{} condition attained the lowest. Additionally, students' attitudes toward LLMs' programming capabilities improved significantly after collaborating with LLM-based tools, and preferences were largely shaped by the perceived usefulness for completing assignments and learning programming skills, as well as the quality of collaboration. Our qualitative findings further reveal that while students appreciated LLM-based tools as valuable pair programming partners, they also identified limitations (e.g., contextual constraints and possibly outdated knowledge bases), and had different expectations compared to human teammates. Students in our study primarily relied on LLM-based tools for syntax clarification and conceptual guidance, while turning to human partners for idea exchanges. Our study provides one of the first empirical evaluations of GenAI as a pair programming collaborator through a comparison of three conditions (\pp, \pai, and \sai). We also discuss the design implications and pedagogical considerations for future GenAI-assisted pair programming approaches.
\end{abstract}



\maketitle

\section{Introduction}
Recent advancements in Generative AI (GenAI) have ushered in a transformative area in programming~\cite{jo2023promise, 10.1145/3490099.3511119, 10176168, 10213396}, driven by breakthroughs in Large Language Models (LLMs). LLM-based applications (e.g., ChatGPT~\cite{ChatGPT}, GitHub Copilot~\cite{copilot}) provide capabilities such as code generating~\cite{10.1145/3544548.3580919, 10.1145/3697010, 10698405} and debugging assistance~\cite{10698405, kang2025explainable}, delivered through forms like chatbots~\cite{dam2024complete, ChatGPT} and inline completions~\cite{copilot, husein2024large}. LLM-based innovations with human-AI collaboration potentials are reshaping programming workflows and posing new challenges in computer science (CS) education~\cite{10.1145/3626252.3630909, 10.1145/3632620.3671116, 10.1145/3568812.3603476}, including traditional human-human collaborative approaches such as pair programming.

Pair programming~\cite{beck2000extreme} has long been established as a powerful collaborative strategy in both CS education and industry practices~\cite{hannay2009effectiveness, hanks2011pair}, recognized for its effectiveness in enhancing problem-solving skills~\cite{10.1145/364447.364614} and code quality~\cite{10.1145/1062455.1062545}. Traditionally, pair programming has relied on the dynamic interplay between two human collaborators, where one performs coding and implementations while the other provides assistance like debugging and reviewing, each fulfilling distinct roles that complement the other~\cite{hannay2009effectiveness, hanks2011pair}. However, the rapid advancement of GenAI introduces a new paradigm shift from an auxiliary tool to a collaborator, providing insights comparable to those of a human partner in pair programming.

As such, understanding GenAI’s role in pair programming and its impact on collaboration in education is key to assessing its effectiveness as a learning and programming partner, which might further shape its potential to enhance human interaction in CS education. While prior research highlights the benefits of traditional pair programming~\cite{10.1145/364447.364614, 10.1145/1062455.1062545} and explores GenAI-assisted coding~\cite{lyu2024evaluating, 10.1145/3639474.3640076}, the prospect of treating LLM-based tools as collaborators rather than mere aids remains largely unexamined~\cite{ma2023ai}. The underexplored potential of GenAI to fully engage in the collaborative dynamics of pair programming underscores the need for classroom deployments that assess GenAI's ability to either complement or replace human programmers. Furthermore, understanding how the integration of GenAI into pair programming processes might influence students' attitudes towards GenAI is crucial as it can inform effective collaboration strategies and enhance overall educational outcomes. 

The following research questions (RQs) guide our work: \\
\textbf{RQ1.} In what ways do collaborations with LLM-based tools as pair programming partners influence students' attitudes towards GenAI on programming? \\
\textbf{RQ2.} How does integrating LLM-based tools as collaborators in pair programming affect student learning outcomes?  and \\
\textbf{RQ3.} What differences emerge in students' experiences and preferences when comparing pair programming with LLM-based tools versus traditional human partners?

To answer these questions, we conducted a mixed-method study from September to December 2024 with 39 students in an undergraduate CS course. Throughout the semester, participants completed six in-class assignments under three pair programming conditions: \textit{Traditional \textbf{P}air \textbf{P}rogramming} (\pp), where two human programmers collaborate without GenAI; \textit{\textbf{P}air Programming with Gen\textbf{AI}} (\pai), in which two human programmers work alongside LLM-based collaborators; and \textit{\textbf{S}olo Programming with Gen\textbf{AI}} (\sai), where a single programmer is paired with LLM-based collaborators. We collected assignment scores and surveyed students' attitudes towards LLM-based tools in the context of pair programming (including both structured survey questions and open-ended responses), as well as students' reflection reports on their collaborations with LLM-based tools.

Our quantitative results show 
1) students' overall attitudes towards LLM-based tools in programming improved significantly after working with LLM-based tools as pair programming collaborators over time, 
2) students' assignment score was highest in the \pai{} and lowest in the \sai{} condition, and 
3) both the collaborative experience and the perceived usefulness influenced students' preferences for the different pair programming conditions.

Our qualitative analysis further revealed that 
1) in general, students considered LLM-based tools as useful collaborators in pair programming, though they noted certain limitations, such as contextual constraints and outdated knowledge bases, 
2) students expressed varying expectations regarding the roles of human collaborators versus LLM-based collaborators during pair programming, where treating LLM-based collaborators more as complements for syntax and concepts assistance while seeking idea exchange with human collaborators, and 
3) students demonstrated different preferences for interaction modalities, with some favoring chatbots and others preferring inline completions as their LLM-based collaborators, creating diverse workflows when working with LLM-based tools in pair programming.

In this work, we contribute: 
\textbf{1)} One of the first empirical studies on pair programming with LLM-based tools as collaborators in a real-world classroom setting by comparing three pair programming conditions, providing empirical evidence that LLM-based tools have the potential to serve as active collaborators during pair programming;
\textbf{2)} An in-depth analysis of human-AI-interaction dynamics in pair programming contexts, revealing how GenAI influences team interactions, learning outcomes, and coding performance as a facilitator or a potential impediment; and 
\textbf{3)} New insights and design implications for future AI-assisted pair programming tools and pedagogical strategies for integrating GenAI into CS education.

\section{Related Work}
\subsection{GenAI in Computer Science Education}
Since the public release of ChatGPT in 2022~\cite{ChatGPT}, GenAI applications have evolved into experimental components of educational technology~\cite{lo2023impact, adeshola2024opportunities,yue2024mathvc}. GenAI demonstrated strong capabilities in coding tasks~\cite{10.5555/3666122.3667065, 10196869}, prompting researchers to investigate its role in CS education~\cite{10.1145/3626252.3630909, 10.1145/3632620.3671116, 10.1145/3568812.3603476}. Recent studies have explored various applications of LLM-based tools in computer science learning contexts, such as automated code explanations~\cite{macneil2023experiences}, enhanced debugging support~\cite{leinonen2023using, 10698405, kang2025explainable}, and generating exercises to help students practice~\cite{sarsa2022automatic, 10.1145/3544548.3580919, 10.1145/3697010, 10698405}, as well as examining how students and instructors perceive  GenAI in educational contexts~\cite{10305701, 10.1145/3568813.3600138, 10343467}.

Despite the growing studies in researching the capabilities and applications of GenAI in CS education, most existing research positions LLM-based tools as supplemental roles, such as acting as teaching assistants~\cite{lyu2024evaluating, mehta2023can, 10.1145/3649217.3653574} or learning companions~\cite{teng2024chatgpt, punar2024cultivating}, and source of learning materials~\cite{10.1145/3639474.3640076, silva2024chatgpt} in entry-level courses~\cite{scholl2024novice, mikac2024chatgpt, scholl2024analyzing}. For example, Lyu et al.~\cite{lyu2024evaluating} explored the utilization of LLM-based tools as teaching assistants in an entry programming course, finding that access to LLM-based tools and human teaching assistants simultaneously enhances students' learning outcomes. Xue et al.~\cite{10.1145/3639474.3640076} conducted a controlled experiment on comparing ChatGPT with traditional search engines as sources of learning materials in the CS1 class environment and found that learning from ChatGPT cannot guarantee enhanced performance immediately. 

However, the idea of treating LLM-based tools as \textit{collaborators}, rather than mere assistants, remains underexamined, especially in contexts beyond introductory courses. Investigating the collaborative potential of GenAI is critical for refining how it is integrated into CS education. By exploring scenarios where LLM-based tools can actively partner with students, we may better understand how GenAI could enhance deeper learning, improve skill development, and reshape longstanding collaborative practices such as pair programming. Our work seeks to address this gap by examining the role of LLM-based tools as collaborators in an undergraduate classroom environment.

\subsection{Pair Programming}
Pair programming is a collaborative software development approach where two programmers work together at a single workstation~\cite{hanks2011pair, hannay2009effectiveness, beck2000extreme}. 
In a pair programming context, one person actively writes the code as the role of ``driver'', while the other provides suggestions with debugging and code reviewing as ``navigator''. The two programmers frequently switch roles to maintain engagement and share responsibility. 
The usage of pair programming in educational settings has increased significantly since the 1990s~\cite{beck2000extreme} and demonstrated various benefits such as increased assignment scores, code quality, confidence, and problem-solving skills~\cite{hanks2011pair,hannay2009effectiveness}. Beyond general advantages, specific pairing strategies significantly shape learning outcomes~\cite{lui2006pair}. Expert-novice pairings can facilitate mentorship and knowledge transfer, though they risk the novice becoming passive if not managed carefully. Conversely, novice-novice pairs may foster mutual learning and confidence but might require additional guidance to develop effective programming practices.

While traditional pair programming underscores the dynamics of collaborations between two human peers, the rise of GenAI has also enabled human-AI collaboration, introducing a new dimension of pair programming~\cite{ma2023ai}. Inline completion tools, such as GitHub Copilot~\cite{copilot}, allow human programmers to work alongside LLM-based tools that can suggest code snippets, identify errors, and provide real-time feedback. Recent studies have begun to explore interaction modes of human-AI pair programming~\cite{10.1145/3586030, 10.1145/3491101.3519665, yan2025llm} and related performance outcomes~\cite{10.1145/3510454.3522684, 10.1145/3544548.3580919, 10.1145/3491101.3519665} from case studies to experimental studies. For example, Barke et al.~\cite{10.1145/3586030} examined the collaboration between human programmers and GitHub Copilot for given tasks. Their studies have identified the two modes of human-AI interactions in coding-getting minor code completions as accelerations and seeking major support as explorations without examining the performance outcomes. In educational contexts, Kazemitabaar et al.~\cite{10.1145/3544548.3580919} conducted a controlled experiment with 69 novice programmers (ages 10–17) using OpenAI Codex for 45 Python tasks. Their findings showed that AI-assisted learners completed tasks 15\% faster and produced higher-quality code and that AI code generators can reduce frustration to novices. 

However, little work has explored the \textit{collaboration dynamics} of pair programming. Collaboration dynamics, a core component of pair programming, refers to the interplay of communication, role distribution, and decision-making processes between partners as they work together on programming tasks~\cite{tan2024collaborative}. These dynamics can impact both the immediate productivity and the long-term learning outcomes of involved members ~\cite{hannay2009effectiveness}. To address this gap, our work explores LLM-based tools specifically in a \emph{pair programming} setting to investigate how GenAI influences collaboration, learning, and performance, compared to traditional human-human collaboration and GenAI-assisted programming. By situating the study in an authentic classroom and examining various pair programming conditions, we aim to enrich our understanding of how GenAI and humans can function together most effectively, thus contributing to human-AI interactions in pair programming at large.

\section{Method}
Upon approval from our Institutional Review Board (IRB), we conducted a semester-long study with 39 students from September (after the course add/drop period) to December 2024 (final exam date). Below, we describe an overview of our participants, our study procedures, and our quantitative and qualitative data analysis.

\subsection{Participants}
\label{subsec:participants}
Our study took place in the Computer Science Department of a four-year university in the United States, within an advanced course primarily focused on teaching \textit{Web Development}. The course syllabus covered HTML, CSS, JavaScript, and related frameworks such as React. To participate, students were required to be 18 years or older, proficient in English for both speaking and writing, and enrolled in an advanced-level undergraduate computer science course.

Of the 39 participants, the majority (38) were majoring in Computer Science, with one in Data Science. One student was a sophomore, eight were juniors, and 30 were seniors. Most participants had prior experience with at least one LLM-based tool, with 37 having used ChatGPT and 12 having used GitHub Copilot, while two students had never used any LLM-based tools before. On average, participants had 4.26 years of programming experience. However, prior exposure to pair programming was less common, with only 6 participants having previous experience, while 33 had never participated in a pair programming setting before. A detailed breakdown of participant demographics is available via \href{https://osf.io/6stpb/?view_only=ea2580afc1a043fe93fd8e275327e0d6}{OSF}.

\subsection{Study Procedure}
\label{subsec:study_procedure}
\begin{figure*}[h!]
\centering
    \Description{Timeline of our study.}
    \includegraphics[width=.82\linewidth]{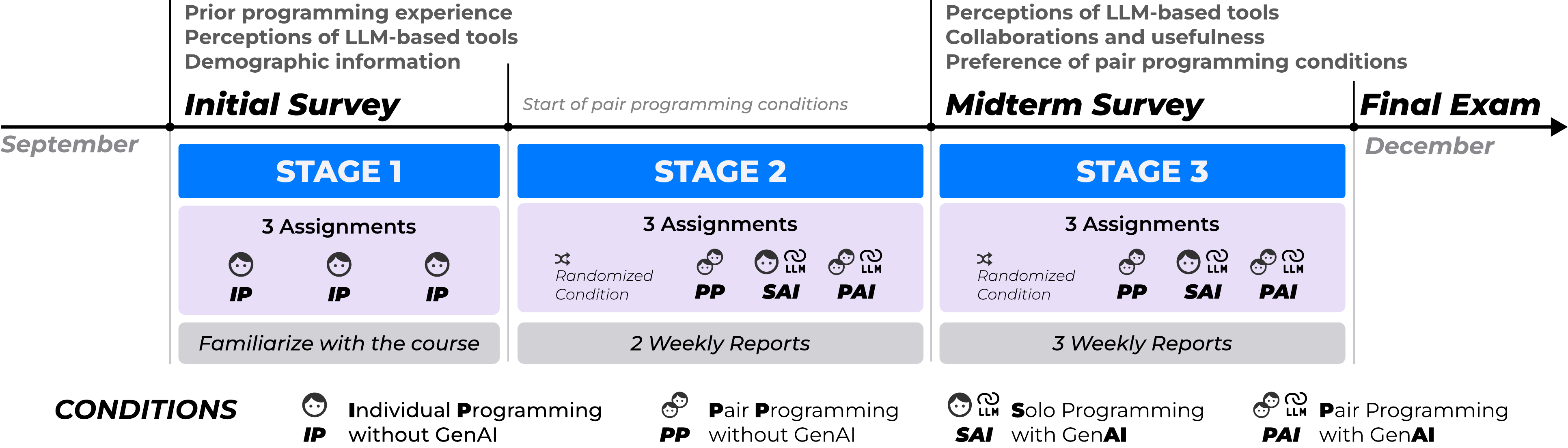}
    \caption{Overall structure of our study. Students completed three assignments in both Stage 2 and Stage 3, each under a different pair programming condition in random order. The final exam reused all questions in the midterm survey.}
    \label{fig:timeline} 
\end{figure*}

Our study procedure includes an initial survey, three stages of course with a midterm survey and a final survey. An overall timeline is shown in \autoref{fig:timeline}. Detailed survey questions and data analysis code can also be found via \href{https://osf.io/6stpb/?view_only=ea2580afc1a043fe93fd8e275327e0d6}{OSF}.

\subsubsection{Initial Survey}
The initial survey included three sections. Section 1 was to understand students' prior experience in general programming and pair programming (e.g., the number of years of programming experience they have had and prior experience in pair programming sessions). Section 2 explored participants' prior experiences with and attitudes toward LLM-based tools. We first asked, \textit{``Which of the following LLM-based tools have you used in the past?''}. For respondents who did not select \textit{``I have no experience using LLM-based tools''}, we further asked about their perspectives and usage patterns, such as \textit{``Please indicate to what extent you agree with the following statements regarding LLM-based tools in programming.''} Section 3 asked about participants' demographic information. 

\subsubsection{Course Structure}
The course followed a flipped classroom structure~\cite{akccayir2018flipped}, where students were expected to study the provided instructional materials before class and engage in interactive activities during class time. In-class activities included programming assignments and discussions with the instructor and classmates. Specifically, there were six pair programming assignments, each due by the end of the respective class session. On non-programming assignment days, students participated in discussions. Additionally, \textit{weekly reports} were required only for the weeks that included a pair programming assignment, where students documented their experiences with the corresponding pair programming tasks after class. The course was structured into three distinct stages temporally:

\textbf{Stage 1, Weeks 1-5.} The first stage of the course was designed for students to become familiar with the course structure. Students were required to form final collaborative project groups on their own. During this stage, students completed three non-pair-programming assignments (Assignments 1-3), with all expected to be completed in the form of \textit{\textbf{I}ndividual \textbf{P}rogramming without GenAI} (\ip) to familiarize themselves with the workflow of assignments. 
Students were then guided through the activation of the GitHub Student Developer Pack, with access to premium GitHub Copilot features, such as unlimited inline code completions. In parallel, we provided a custom web-based platform developed by the research team, which allowed students to access OpenAI’s paid models, features typically restricted to subscribed users. A tutorial lecture was also offered to help students integrate these LLM-based tools into their programming workflow. Students were strongly encouraged to use our web-based tool exclusively for all conversational interactions with LLM-based tools throughout the course.

\textbf{Stage 2, Weeks 6-9.} During the second stage, students completed three assignments (Assignments 4–6) under three different pair programming conditions: \pp, \pai, and \sai. Each student was randomly paired with one or two teammates from their final project group. The three conditions were assigned in a randomized order, with each student experiencing each condition once across the three assignments.
Starting from Stage 2, all assignments were graded using predefined autograders, which evaluated the correctness of the submitted code based on its running output. At the end of Stage 2, weekly report 6 was distributed in a midterm survey format to gather students' perspectives on pair programming and LLM-based tools. The survey included questions from the initial survey, such as \textit{``Please indicate to what extent you agree with the following statements regarding LLM-based tools in programming''}. Additionally, new questions were introduced to assess students’ experiences collaborating with both LLM-based tools and human teammates, (e.g., \textit{``Please indicate to what extent you agree with the following statements about your collaboration with LLM-based tools/your human pair programming teammate(s) in the three assignments''}. Additionally, students were asked to rank the four programming conditions (\pp, \pai, \sai, and \ip) from most to least preferred and provide their perceptions of the usefulness of both pair programming and LLM-based tools in completing assignments and improving their programming skills.

\textbf{Stage 3, Weeks 10-15.} In the third stage, students completed three additional assignments (Assignments 7–9) while working with the same human teammate(s) throughout. Similar to Stage 2, students experienced each of the three pair programming conditions (\pp, \pai, and \sai) once, with the order of conditions randomized across the three assignments. 

\textbf{Final Exam.} The final exam was conducted in person during the last week of the semester and focused on gathering students' opinions on pair programming and LLM-based tools. All questions from the midterm survey were included in the final exam, allowing for a direct comparison of responses.

\subsection{Data Collection and Analysis}
\label{subsec:data_analysis}
\subsubsection{Quantitative Data Collection \& Analysis}
The quantitative data collected includes scores of students' programming assignments and structured data from surveys with students’ attitudes toward LLM-based tools, perceptions of pair programming conditions, and future programming preferences. We analyzed quantitative data with multiple statistical methods. In all statistical analyses, we set the significance level at $p < 0.05$.

For quantitative analysis, we first computed descriptive statistics for all variables, reporting means and medians for continuous variables. To assess changes in students’ attitudes toward LLM-based tools over time, we conducted a \textit{Friedman test} (using R package stats ~\cite{stats}) to examine differences in agreement levels across three time points: Initial, Midterm, and Final. When significant differences were detected, we performed \textit{Wilcoxon signed-rank tests} (using R package stats~\cite{stats}) for post-hoc pairwise comparisons, adjusting p-values to account for multiple comparisons.

To examine students’ performance under different programming conditions, we analyzed scores of six pair programming assignments (Assignments 4 to 9). Given the non-normality of score distributions, we employed a \textit{Kruskal-Wallis test} (using R package stats ~\cite{stats}) to determine whether significant differences existed among the instructional setups. When a significant effect was found, we conducted post-hoc pairwise comparisons using \textit{Dunn’s test} (using R package FSA~\cite{fsa}) to identify which groups differed.

To further investigate factors influencing students’ future programming preferences, we applied \textit{Cumulative Link Mixed Models} (using R package ordinal ~\cite{ordinal}) to examine the impact of students’ perceptions of GenAI and pair programming on their likelihood of choosing \ip, \sai, \pp, or \pai. The model included predictors such as perceived usefulness and collaboration for both GenAI and pair programming, as well as time as a factor.

\subsubsection{Qualitative Data Collection \& Analysis}
The qualitative data produced includes students’ weekly report assignments and open-ended responses from surveys.

We used the \textit{General Inductive Approach}\cite{thomas2006general} for this analysis. Initially, the first author conducted a close reading of the entire dataset to identify recurring patterns and key ideas. Relevant text segments were then systematically labeled to generate preliminary categories, forming the basis for developing low-level codes that captured the nuances of students’ perspectives. These low-level codes were subsequently aggregated into high-level themes representing overarching concepts in the data, such as \textit{efficiency}, \textit{skill Levels}, and \textit{workflow}. Throughout the coding process, the research team engaged in iterative discussions to refine and validate emerging themes. A codebook outlining the themes and codes is provided via our \href{https://osf.io/6stpb/?view_only=ea2580afc1a043fe93fd8e275327e0d6}{OSF} materials.

\begin{figure*}[h!]
\centering
    \Description{An overview of students' average attitudes across five opinion statements about programming with LLM-based tools, from the beginning to the midterm and final stages.}
    \includegraphics[width=.78\linewidth]{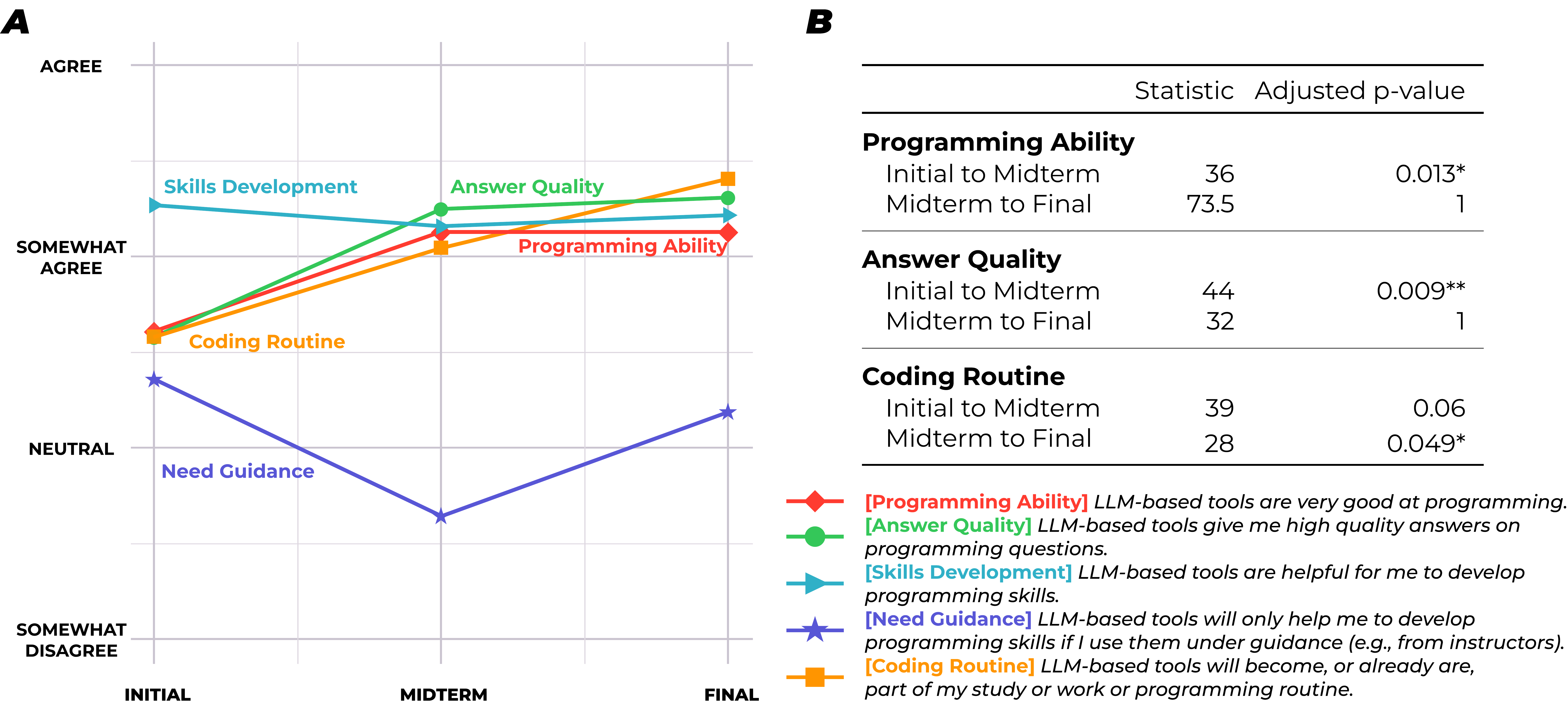}
    \caption{(A) An overview of students' average attitudes across five different opinion statements about programming with LLM-based tools, tracking changes from the initial to the midterm and final stages; (B) Non-parametric pairwise comparison test (Wilcoxon test): Differences in agreement levels across various opinion statements about programming with LLM-based tools. Only statements with significant changes are included. (Significance level: * $p < 0.05$, ** $p < 0.01$).}
    \vspace{-\baselineskip}
    \label{fig:attitude} 
\end{figure*} 
\section{Quantitative Results}

In this section, we first report students' attitudes toward LLM-based tools in programming and the trends of their attitudes change over time. We then examine students' performance differences in assignments among three pair programming conditions, as well as their preference for programming conditions in the future.

\subsection{Students' Attitudes towards LLM-based tools on Programming}
\label{subsec:attitude}

To evaluate how students’ attitudes toward LLM-based tools for programming evolved over the semester, we measured students' agreement with five statements at three different time points (initial, midterm, and final), with the significant results of Friedman tests, as shown in \autoref{fig:attitude}. 

For the statement related to \textbf{Coding Routine} (\textit{LLM-based tools will become, or already are, part of my study/work/programming routine}), we see a steady and continuous increase in agreement over time. Although the change from Initial to Midterm was not statistically significant ($p = 0.06$), the increase from Midterm to Final reached significance ($p < 0.05$), indicating that as students collaborated more with LLM-based tools, they increasingly accepted these tools as an integral component of their programming workflows. 

Similarly, the statements assessing LLM-based tools' \textbf{Programming Ability} (\textit{LLM-based tools are very good at programming}) and \textbf{Answer Quality} (\textit{LLM-based tools give me high-quality answers on programming questions}) showed a significant rise from the initial stage to the midterm ($p < 0.05$ or lower), followed by stabilized yet sustained increase by the final stage, indicating the potential impact of early exposure on students' evaluations of based tools’ capabilities and output quality in programming. In contrast, the perception of LLM-based tools as beneficial for \textbf{Skills Development} (\textit{LLM-based tools are helpful for me to develop programming skills})  remained relatively stable throughout the semester, with only minor fluctuations observed, implying a relatively stable perception of LLMs' role in programming skill development over time. The essence of \textbf{Need Guidance} (\textit{LLM-based tools will only help me to develop programming skills if I use them under guidance (e.g., from instructors)} demonstrated a more dynamic trend. After an initial decline before the midterm, there was a progressive increase in agreement toward the end of the semester. 

Overall, our findings reveal that students' \textbf{attitudes toward LLM-based tools grew more positive over time} in the context of pair programming, both regarding their programming abilities and the idea of incorporating them into regular coding routines. Although \textbf{early positive interactions with LLM-based tools foster students’ confidence and promote routine use in programming}, the subsequent \textbf{shift in the perceived need for guidance, from an initial decline to a progressive rise,} suggests that structured external support might be crucial to programming skills development.

\subsection{Assignment Performance \& Future Adoption}
\label{subsec:performance}

\subsubsection{Comparative Analysis of Scores}

We computed the average scores for each setup group across Assignments 4 to 9. The \pai{} condition exhibited relatively stable performance, with scores ranging from 41.33 to 60. In contrast, the \pp{} condition showed greater fluctuations, with scores varying between 31.11 and 66.75. Similarly, the \sai{} condition demonstrated variability, with scores spanning from 28.21 to 48.57. Through descriptive analysis of the mean scores for each setup across assignments, we observe variations in score distributions among instructional setups. To further examine these differences and minimize the risk of short-term fluctuations caused by topic difficulty, external factors, or one-time anomalies, we below analyze the full dataset rather than evaluating assignments individually.

To assess whether significant differences exist between groups, we conducted a Kruskal-Wallis test after confirming that score distributions across setups deviate from normality. The results revealed a significant effect, with $\chi^2 = 6.69$ ($p < 0.05$), indicating that at least one setup differs significantly from the others. Examining the median scores, we observe that students in the \pai{} achieved the highest median ($M=60$), followed by \pp{} ($M=47.14$) and \sai{} ($M=33.33$), suggesting that instructional setup may influence student performance.

\begin{figure}[h!]
\centering
    \Description{Non-parametric pairwise comparison test (Dunn’s test): Differences in assignment scores across different groups. Students achieved the highest scores when working in \pai{} conditions, followed by \pp, while the lowest in \sai.}
    \includegraphics[width=.8\linewidth]{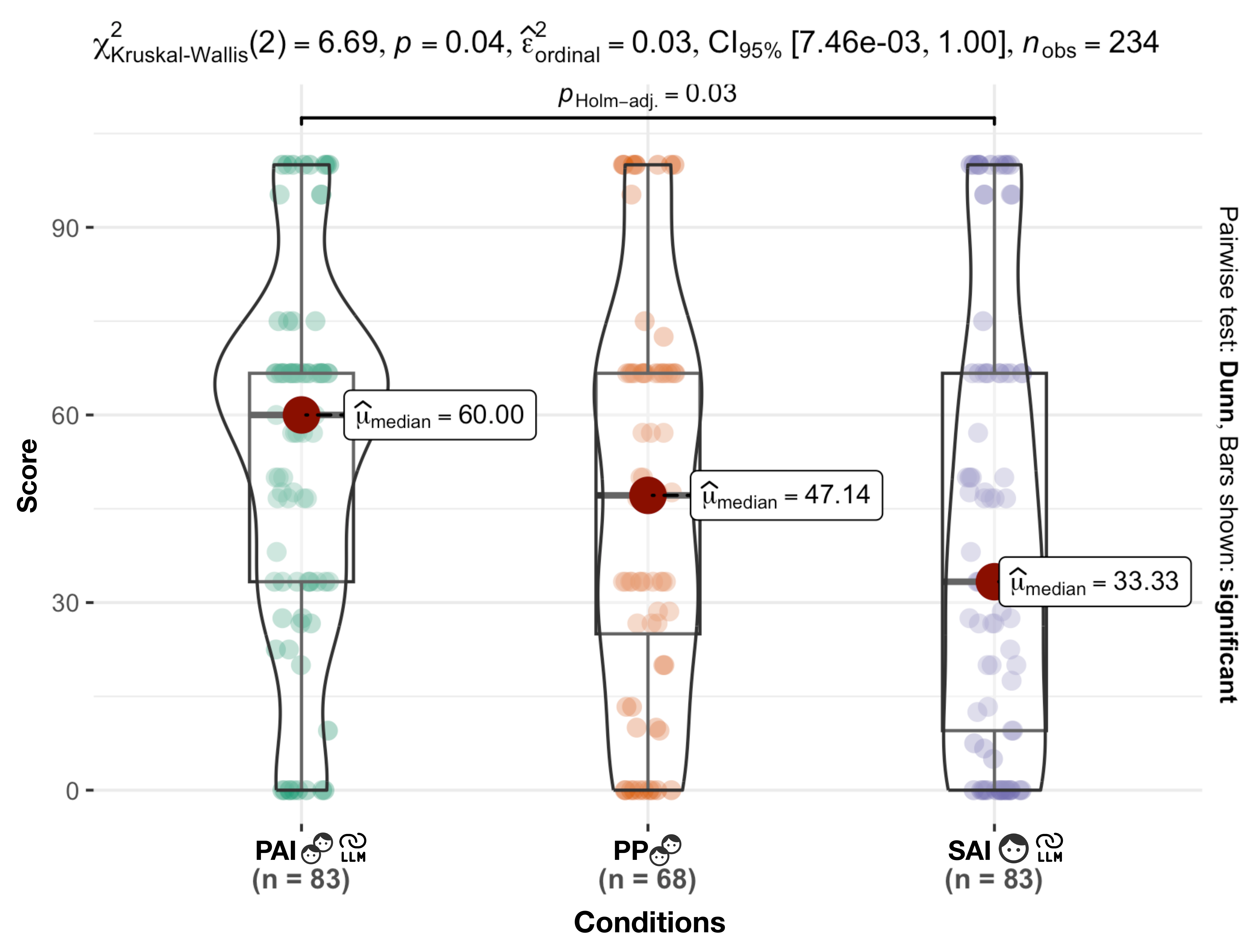}
    \caption{Non-parametric pairwise comparison test (Dunn’s test): Differences in assignment scores across different groups. Students achieved the highest scores in \pai{} conditions, followed by \pp, while the lowest in \sai.}
    \vspace{-\baselineskip}
    \label{fig:grade} 
\end{figure} 

To further investigate where these differences occur, we conducted post hoc pairwise comparisons using Dunn’s test (see \autoref{fig:grade}). The results revealed a significant difference between the \pai{} and \sai{} ($p < 0.05$) that suggests \textbf{students in the \pai{} condition consistently outperformed those in the \sai{} condition, indicating the benefits of working with a human partner and LLM-based tools simultaneously in completing educational programming tasks.}

\subsubsection{Exploring Factors Influencing Students' Future Usage Decisions}

\begin{table*}[h!]
    \small
    \centering
    \caption{Cumulative Link Mixed Models explaining respondents' \textit{preference} in how they would like to continue programming in the future (Significance level: * $p < 0.05$, ** $p < 0.01$, *** $p < 0.001$).}
    \label{tab:cumulative_link_mixed_models}
    \footnotesize
    \begin{tabular}{lllll}
    \toprule
     & \multicolumn{4}{c}{\textbf{Dependent Variable}} \\
    \cline{2-5}
     & SP  &SAI & PP & PAI \\
     & $\beta$ (Std. Error) & $\beta$ (Std. Error) & $\beta$ (Std. Error)& $\beta$ (Std. Error)\\
    \midrule
    Time & 0.929 (0.624) &0.968 (0.603)&-0.810 (0.575)&-0.787 (0.590)\\ 
    \cdashline{1-5}
    \textbf{Usefulness} &&&\\
    Pair Programming Usefulness in Assignments & 0.261 (0.300) & -0.972 (0.396)&0.576 (0.316) &0.144 (0.301)\\
    Pair Programming Usefulness for Programming Skills & -0.970 (0.307)** & -0.257 (0.299)&0.503 (0.293) & 0.851 (0.333)*\\
    GenAI Usefulness in Assignments & -0.706 (0.332) &0.337 (0.443) &0.479 (0.389) & -0.025 (0.390) \\
    GenAI Usefulness for Programming Skills & -0.179 (0.350) &-0.715 (0.408) &0.180 (0.344) & 0.441 (0.390)\\
    \cdashline{1-5}
    \textbf{Collaboration} &&&\\
    GenAI Collaboration & -0.806 (0.350) &2.319 (0.735)* &-1.936 (0.610)** & 0.719 (0.616)\\
    Pair Programming Collaboration & -1.126 (0.479)* &-0.805 (0.518)&0.837 (0.430)&0.706 (0.442)\\   
    \cdashline{1-5}
    \textbf{Const}&&&\\
    Forth Choice | Third Choice  & -16.258 (3.131)*** &-4.535 (2.720) &1.183 (2.063)& 8.806 (2.757)**\\
    Third Choice | Second Choice  & -15.345 (3.047)*** &-1.855 (2.571) &4.754 (2.305)& 11.833 (3.126)***\\     
    Second Choice | First Choice  &  -12.695 (2.788)*** &0.824 (2.471)&8.090 (2.470)**&13.991 (3.419)***\\
    \bottomrule
    \end{tabular}
\end{table*}

\autoref{tab:cumulative_link_mixed_models} presents the results of the Cumulative Link Mixed Models, explaining respondents' \textit{preference} for different programming conditions: \ip, \sai, \pp, \pai. The predictors include perceptions of usefulness and collaboration regarding both pair programming and LLM-based tools, as well as the effect of time.

For \ip, the perception that pair programming improves programming skills significantly reduces the likelihood of selecting this condition ($\beta = -0.970$, $p < 0.01$). Additionally, those who find pair programming beneficial for collaboration are significantly less likely to prefer working alone ($\beta = -1.126$, $p < 0.05$). Meanwhile, perceptions of LLM-based tools' usefulness, collaboration, and time spent programming do not significantly influence students' choices.

For \sai, a key positive predictor is a favorable attitude toward working with LLM-based tools, which significantly increases the preference for this condition ($\beta = 2.319$, $p < 0.05$). This suggests that individuals who view LLM-based tools as a strong collaborative tool are more inclined to choose GenAI-assisted solo programming. However, for \pp, a negative attitude toward LLM-based tools' collaboration significantly increases the preference for traditional pair programming ($\beta = -1.936$, $p < 0.01$), indicating that respondents who do not perceive LLM-based tools as effective collaborators may favor traditional pair programming instead.

For \pai, the strongest predictor is the perception that pair programming enhances programming skills, which significantly increases the likelihood of preferring this condition ($\beta = 0.851$, $p < 0.05$). However, perceptions related to LLM-based tools, including their usefulness for assignments and programming skills, do not show significant effects, indicating that the assistance of LLM-based tools alone does not necessarily drive preference for this condition.

\takeaway{Overall, our quantitative results suggest that pairing with both LLM-based tools and a human partner (\pai) provided the greatest benefit for student performance, outperforming both traditional pair programming (\pp) and solo programming with GenAI (\sai). Students’ future preferences for pair programming setups were strongly tied to their perceptions of collaboration and tool effectiveness: those who did not view LLM-based tools as effective collaborators preferred traditional pair programming, whereas a positive perception of GenAI’s usefulness increased the likelihood of choosing \sai. Moreover, believing that pair programming itself enhances programming skills significantly raised the preference for \pai.}
\vspace{-\baselineskip}

\section{Qualitative Findings}
Our qualitative analysis of students' weekly reports and final exams offers additional insights into the similarities and differences among three pair programming conditions. Below, we explore how students navigated the evolving roles of LLM-based tools and human peers, as well as the strategies they used for collaboration.

\subsection{Team Dynamics: Navigating Collaboration with Human and AI Partners}
\label{subsec:team_dynamic}
\subsubsection{Balancing Skill Levels in Pair Programming}
Students in our study \textbf{expressed different preferences for the skill levels of human teammates}, despite traditional educational settings often considering pairing students of similar skill levels as the most effective collaboration in pair programming~\cite{chaparro2005factors}. For example, students found pairing two inexperienced students can sometimes be counterproductive. Without external guidance, inexperienced teams often found themselves stuck in problem-solving deadlocks. As P23 reflected in such situations:

\quoteblock{``In general, if I didn't know how to do something, my teammate would also not know how to do/code it, ... we would just end up having to ask the professor and the TA.''}{P23}

P23's quote states that pairing inexperienced students together may lead to increased confusion and anxiety. When both partners lacked foundational knowledge, students' progress was significantly hindered, often resulting in frustration and inefficiencies. Moreover, the frequent need for external intervention disrupted the collaborative experience, potentially diminishing students' confidence and autonomy in problem-solving.

Conversely, some students found that pairing individuals with different skill levels, similar to industry practices, produced mixed outcomes. In some cases, less experienced students benefited from working with a more skilled partner as expected:

\quoteblock{``He [human partner] was far more experienced ... and it really helped me to be able to learn from him and work with him.''}{P7}

P7's experience highlighted the advantages of working with a more experienced partner in pair programming. By observing a skilled teammate’s approach to problem-solving and implementation, less experienced students could accelerate their learning through hands-on exposure. However, the knowledge gap between teammates does not always lead to positive outcomes, as P2 stated:

\quoteblock{``My teammate's computing skill is much better than [mine]. In this case, there are gaps between us. Sometimes it is hard for me to catch up.''}{P2}

P2’s quote illustrates the challenges that can arise from a significant skill gap between teammates in pair programming. When one partner is considerably more experienced, the less skilled individual may struggle to keep up, leading to confusion and limiting their ability to effectively learn from the process. Meanwhile, disparities can also make advanced students feel inefficient, as they may need to slow down or take on a disproportionate share of the work:

\quoteblock{``There were times when the knowledge discrepancy between me and my partner felt like a lot. So much so that it felt like I was almost doing an individual assignment. They definitely got better as the semester went on and they learned. However, during those early assignments I felt like pair programming was getting in my way because all it was doing was interrupting my traditional problem solving routine.''}{P17}

P17’s statement highlights another drawback of a significant skill gap in pair programming—while experienced students are not necessarily opposed to working with less-skilled partners, they can find the experience frustrating. As P20 also stated, \emph{``there was also some stress caused by watching your partner work slowly.''} For more experienced students, the added responsibility of guiding their teammates can feel burdensome, making the process less efficient and even disruptive to their usual problem-solving approach. Unlike in industry settings, where mentoring junior engineers is an expected part of the workflow, the primary goal of pair programming in an educational context is to complete assignments together. As a result, more experienced students may perceive collaboration as an obstacle in pair programming, which could increase cognitive overhead and interrupt their workflow.

\subsubsection{LLM as a Bridge: Mitigating Skill Gaps}
\textbf{The integration of LLM-based tools into pair programming can benefit inexperienced students in navigating challenges more effectively.} P7, who had struggled when paired with another novice student, emphasized:

\quoteblock{``If you pair up two people that have never had any experience ... please please please, let them have help from GenAI. When I was paired with another person who had no prior experience, we did not have help from GenAI and we really struggled when we ran into issues... led to poor learning, high levels of stress, and bad grades.''}{P7}

P7's reflection underscores the value of incorporating LLM-based tools into pair programming for inexperienced pairs. By providing guidance on where and how to start, LLM-based tools can alleviate stress and anxiety while preserving the teamwork dynamic. The intervention of LLMs offers a structured approach to problem-solving, helping novice students overcome obstacles more efficiently and fostering a more positive and productive learning experience.

As for mixed-skill pairs, LLM-based tools facilitated more balanced participation. Instead of the more experienced partner dominating the entire process, integrating LLM-based tools into pair programming enabled both students to remain actively engaged. P33 described one such situation:

\quoteblock{``My partner was the one to use the chat AI, I used GitHub Copilot. The Copilot was helpful ... all I had to do was look it over and ensure the functionality worked and run the tests.''}{P33}

As the more experienced student in their pair, P33 was able to focus on coding with the support of GenAI, allowing them to fully engage in the pair programming process. Meanwhile, P33's less experienced teammate could utilize conversational LLM-based tools to grasp fundamental concepts and gain assistance in understanding implementations. Such a dynamic enabled both partners to find meaningful roles in their collaboration, ensuring that skill level differences did not hinder participation but instead fostered a more inclusive and productive learning experience in pair programming.

\subsubsection{Distinct Expectations of Human \& AI}
As the collaboration process progressed, \textbf{students gradually developed distinct expectations for their human pair programming partners and LLM-based tools}. LLM-based tools were primarily perceived as technical assistants, offering syntax-related support, debugging assistance, and quick explanations of basic concepts. As P3 noted:

\quoteblock{``Something I have discovered is that, when looking for syntactic explanations, GenAI is super helpful.''}{P3}

P9 also pointed out that LLM-based tools remained preferable for simple tasks compared to seeking help from their human partner:

\quoteblock{``Working in pairs helped, but there were still times where I wanted to ask AI about things like syntax, since I am still not too familiar with JavaScript.''}{P9}

These reflections highlight students' expectations of LLM-based tools as sources of rapid assistance for programming syntax and fundamental concepts. LLM-based tools allow students to receive instant help, particularly when working with unfamiliar programming languages or frameworks. Additionally, LLM-based tools enable beginners to ask vague or incomplete questions without needing to meticulously structure their queries, unlike traditional search engines, which require more precise phrasing for effective results.

Conversely, students relied on their human partners for idea exchanges and collaborative problem-solving. As P33 noted:

\quoteblock{``I was able to effectively communicate with my teammate, and we were able to work through tough problems or differences ... by discussing the best approach.''}{P33}

P33's quote underscores a key advantage of pair programming, meaning that students can actively exchange ideas and refine their problem-solving strategies together. As P23 also mentioned, \emph{``it felt nice to have a teammate to bounce ideas off of. I definitely felt more efficient working with a teammate.''} Similarly, P39 observed that \emph{``our discussions led to us learning from each other's coding techniques.''} 

These reflections suggest that while LLM-based tools provide quick technical assistance, human interactions foster deeper engagement, richer learning experiences, and mutual skill development. Discussions with a human partner encourage critical thinking, creativity, and the exploration of multiple problem-solving approaches—elements that are crucial for developing a strong programming foundation beyond mere syntax and debugging support.

\subsection{Integrating GenAI into Pair Programming: Preferences, Adaptive Strategies, and Limitations}
\subsubsection{Varied Preferences: Inline Completion v.s. Conversational Tools}
\textbf{Students expressed varied preferences regarding inline completions and conversational LLM-based tools in their pair programming process.} For example, P18 praised inline completions as \emph{``in-line suggestions are honestly some of the best things I've ever used''}. In contrast, P8 described a scenario where neither Copilot’s inline suggestions nor its chat feature resolved an issue, until they switched to ChatGPT, which quickly identified it:

\quoteblock{``I had an issue that I spent almost an hour trying to fix, and neither Copilot [in-line suggestions] nor the chat feature  [in Copilot] knew how to help. Eventually, I put it in ChatGPT and easily identified a space missing.''}{P8}

By semester’s end, P8 noted, \emph{``I tried to avoid accepting large Copilot suggestions, because sometimes they would be incorrect and confusing.''} These insights indicate that while inline completions from LLM-based tools provide immediate, context-sensitive suggestions, conversational LLM-based tools with larger contexts can deliver more detailed and precise responses for complex queries.

\subsubsection{Shifting from Single to Multiple Screens: Adapting to AI Integration}
\label{subsubsection:screens}
\textbf{The existence of LLM-based tools as separate applications from code editors decreased students' collaborative efficiency in pair programming, while students adapted innovative strategies, such as working on multiple screens, to mitigate.} Although incorporating GenAI into pair programming has clear benefits, with students in \pai{} conditions achieving higher median assignment scores compared to \pp{} conditions in \autoref{subsec:performance}, several participants noted that integrating LLM-based tools can sometimes disrupt existing collaborations, such as the need to shift focus from coding to interacting with an LLM-based tool in a separate window demands extra attention. P19 explained:

\quoteblock{``For the pair-programming part, I felt it’s harder to cooperate on coding while using AI, since there will be a more individual conversation with AI that needs instantaneous reaction.''}{P19}

P19’s comment indicates that the current text-focused communication method with LLM-based tools diverts students' attention from coding, where verbal exchanges of ideas in traditional pair programming between teammates allow for simultaneous coding and discussion. Similarly, another student noted:

\quoteblock{`The only real issue with GenAI tools was the use of it was on a separate window than my code editor it takes extra time to use.''}{P33}

P33's observations suggest that the current mainstream modality for interacting with LLM-based tools as a separate application can lead to decreased efficiency in collaborative settings. Even when using GitHub Copilot as an extension within the coding environment, students sometimes still need the chat features for further queries beyond the inline completion, which still requires students to focus on text-based conversation.

In response, students have created adaptive strategies to balance collaboration with both human teammates and LLM-based tools by working on multiple screens instead of the single-screen coding nature of traditional pair programming. For example, P29's group divided responsibilities between teammates by having one partner code while the other researched methods and information:

\quoteblock{``We were able to have one person actively code while the other could support and research corresponding methods and information for the first JavaScript assignment.''}{P29} 

Similarly, P5’s team split tasks between two screens—one for coding and one for looking up information—\emph{``I looked up questions with my laptop and the code was on my teammate's laptop''}, where P33' team used different LLM-based tools concurrently as \emph{``My partner was the one to use the chat AI, I used GitHub Copilot.''} These adaptive methods empowered students to fully exploit the potential of LLM-based tools in pair programming without being limited by the traditional single-screen settings. 

As P8 further noted, using separate devices allowed both teammates to explore different approaches simultaneously, enhancing efficiency, as they are \emph{``both trying different things''}. These quotes illustrate how students proactively optimized their collaborative processes to mitigate the additional challenges of integrating LLM-based tools into pair programming, thereby enhancing both productivity and communication effectiveness.

Therefore, students' need to work simultaneously on both screens is also expressed by their overwhelmingly positive attitudes toward pair programming extensions like Visual Studio Live Share~\cite{liveshare}:

\quoteblock{``...we used the Live Share VSCode extension for the first time and it was incredibly helpful. This made for a more efficient experience and it was easier to iterate and test as I could easily run the test files or look in the browser console as they worked and vice-versa. Especially for the bonus, it was a lot faster to complete with having both of us work simultaneously on different aspects of it.''}{P20}

P20's account suggests that when collaborative tools are well-integrated, they not only enhance efficiency by allowing simultaneous work but also preserve individual workspace preferences. Such tools help mitigate the friction often associated with shared screens, thereby fostering a more productive and satisfying collaborative environment in pair programming.

\subsubsection{Limitations of LLM-based Tools}
\label{subsec:limitations}
Despite the benefits, \textbf{collaborations with LLM-based tools through pair programming context also made students aware of several limitations of LLM-based tools in programming.} One primary challenge is the importance of prompt engineering and contextual limitations when solving complicated tasks, as P15 observed:

\quoteblock{``Sometimes when the prompt is not clear enough, it will get stuck with a wrong answer even after multiple times of fixing the description. In this case, I’ll have to restart the whole conversation for it to get rid of the context of the previous conversation. It’s time consuming to check if it is stuck in this kind of loop.''}{P15}

P15’s experience highlights how ambiguous input can trap LLM-based tools in error loops, consuming valuable time that could be better spent on productive work. P15’s experience also reveals how accumulated conversational deviations can degrade output quality, a concern echoed by P37, \emph{``it's easy for the AI to mix up code and make our previous code very muddled''}, which further illustrates that a loss of contextual continuity in LLM-based tools, when the length of current conversation is over maximum tokens, can severely hinder progress during coding sessions.

Moreover, students expressed concerns about GenAI hallucinations in programming. As P17 stated:

\quoteblock{``[Hallucinations] led to a lot of confusion on my end and I actually wasn’t able to fully complete the assignment... I had to do a quick mental reset and turn off Copilot.''} {P17}

Here, P17’s experience underscores how hallucinations can cause significant delays and stress, forcing students to interrupt their workflow. The hallucinations are even more pronounced when working with rapidly evolving technologies. When tackling the newly released Svelte 5~\cite{svelte} in one assignment, P28 observed:

\quoteblock{``...it was especially hard to use [LLM-based tools] since Svelte had had a recent update where its syntax drastically changed.''}{P28}

P28's quote illustrates the struggle of LLM-based tools in handling rapidly evolving technologies, as recent updates might not be reflected in training data. Similarly, P36 noted, \emph{``Since generative AI cannot be updated with the most up-to-date information immediately, some information it gives can create more confusion, rather than provide clarity.''} Such discrepancies led P17 to conclude that the AI became \emph{``really bad at writing the actual syntax''}, prompting a reconsideration of integrating LLM-based tools into their workflow.

\takeaway{Our qualitative findings illustrate the complex interplay between students and LLM-based collaborators in pair programming. While LLM-based tools offer valuable technical assistance and can help mitigate the challenges posed by skill disparities, their effectiveness is constrained by issues related to prompt clarity, contextual limitations, and rapidly evolving technology. Concurrently, human interaction remains essential for deeper collaborative learning and idea exchange.} 

\section{Discussion}
Our study results shed light on how students perceive and interact with LLM-based tools in pair programming, and provide insights into the collaboration dynamics. Based on our findings, we discuss several design implications to better support pair programming with LLM-based tools, as well as reflections on fostering GenAI literacy in programming.

\subsection{Design Implications for LLM-based Tools in Pair Programming}
Our findings indicate the potential of LLM-based tools as valuable collaborators in pair programming, yet they also reveal some challenges that may limit their effectiveness. 

First, while students praised LLM-based tools for programming capabilities, they also reported issues with contextual limitations, prompt sensitivity, and outdated knowledge bases. These issues suggest that future tools need stronger contextual awareness and memory, ideally integrated directly into the coding environment. Embedding advanced LLM features within IDEs (Integrated Development Environments) (e.g., deeper code context, search for recent information) would reduce time spent re-establishing context or coping with stale knowledge. Tools like Cursor~\cite{Cursor} hint at this next step by offering enhanced code awareness among multiple files, and ChatGPT supports for ``working with code'' mode to provide assistance with up-to-date information. However, more research is needed to assess the effectiveness of those innovations in pair programming through field studies or controlled experiments.

Second, our qualitative results show that students typically relied on LLM-based tools for technical and syntax-related support, while they depended on human teammates for open-ended idea exchange and conceptual discussions. These findings shed light on design implications that LLM-based solutions should differentiate between quick syntax help and deeper, more advisory roles. Perhaps, allowing users to select a ``skill level'' for the AI (e.g., basic, intermediate, advanced) could tailor the learning experience more effectively than current approaches.

Additionally, our findings indicate that traditional human pair programming enables students to discuss ideas without losing focus on their code. Unlike currently available LLM-based tools, which rely heavily on text-based interactions that can disrupt workflow, future pair programming interventions could integrate voice communications directly into the IDE, allowing users to engage in audio interactions that automatically access current code context, supporting seamless conversations similar to human pair programming.

Furthermore, students' preference for working with LLM-based tools in pair programming on multi-screen setups, as illustrated in \autoref{subsubsection:screens}, also echoes the concept of Distributed Pair Programming (DPP), where two programmers may work from different workstations remotely~\cite{10.1007/3-540-45672-4_20}. Integrating multi-screen environments might provide a more effective approach to handling the increased information input and output inherent in human-AI pair programming. By distributing tasks across multiple screens, students can seamlessly manage generated suggestions from GenAI, code implementation, and debugging processes, which may lead to improved workflow efficiency and enhanced collaborative experiences.

Beyond these functional considerations, it is also crucial to consider non-functional requirements~\cite{chung2012non} when designing LLM-based tools for pair programming. Non-functional requirements, such as system responsiveness, usability, and the ability to support collaborative environments effectively, also play an important role in facilitating a seamless interaction between users and technologies~\cite{saroja2023functional}. 
For example, usability can be prioritized to reduce cognitive load and streamline interaction, allowing students to focus on problem-solving rather than tool navigation. Relatedly, enhancing integration across multiple screens and devices can further support efficient collaboration and collective communication with GenAI, so that LLM-based tools better complement, rather than disrupt, the programming workflow.

\subsection{Fostering GenAI Literacy in Programming: When, How, Why, and Ethical Considerations}

Our findings show that students’ attitudes toward LLM-based tools improve significantly with \emph{early exposure}, reflected in the jump between baseline and midterm attitudes toward output quality. However, determining what ``early'' means in practice requires careful consideration. For example, researchers have concerns that introducing LLM-based tools too soon might lead beginners to over-rely on AI-generated solutions before mastering fundamental skills~\cite{zhai2024effects}, while introducing them too late might miss an opportunity to shape positive, informed perceptions of GenAI’s capabilities and limits~\cite{10343467}. A balanced approach could be to start at intermediate-level courses once students have developed basic programming proficiency, accompanied by structured support in using AI tools. Relatedly and crucially, \emph{structured guidance} might be important to leveraging early exposure effectively, as indicated by our results. Educators may consider scaffolding learning~\cite{gibbons2002scaffolding}, such as by showing students how to construct prompts, interpret outputs, and validate solutions rather than accepting them at face value. More work is needed to establish evidence-based guidelines that help educators integrate LLM-based tools at the optimal point in the CS curriculum. 

At the program level, such a balanced introduction raises important \emph{ethical considerations}. If students become heavily dependent on LLM-based suggestions, they risk diminishing the very problem-solving skills that CS education aims to strengthen. Instructors might weigh the convenience of AI assistance against the possibility of undermining students’ autonomy and creativity. An emerging body of work has looked into how to encourage students to validate AI outputs~\cite{lyu2025understanding} and reflect on the origins and reliability of the underlying data can help address these ethical dilemmas~\cite{theophilou2023learning}. Future work can expand to empirically assess the long-term cognitive and ethical impacts of LLM integration in CS education.

\section{Limitations}
Our study has several limitations that suggest avenues for further research. First, the study was conducted on a relatively small scale, which limits the generalizability of our findings. Second, although we investigated the impact of LLM integration over an entire semester, traditional pair programming studies often examine longer-term effects on students, indicating that further research is needed to capture sustained, long-term dynamics. Third, the controlled educational setting of our study may not reflect the complex nature of professional programming environments; future research should explore other real-world contexts. 

\section{Conclusion}
We conducted one of the first empirical studies that compared three human-AI pair programming conditions (\textit{pair programming without GenAI}, \textit{pair programming with GenAI}, and \textit{solo programming with GenAI}) with 39 students in a real undergraduate classroom. Our findings indicate that students’ attitudes towards LLM-based tools improve significantly during the early stages of exploration. The highest assignment performance was observed under the \textit{pair programming with GenAI} condition, while the lowest occurred in the \textit{solo programming with GenAI} condition. In pair programming collaborations, students primarily sought fundamental technical assistance from LLM-based tools, yet they valued human partners for exchanging ideas and tackling complex challenges. Students also expressed distinct preferences for different modalities of LLM-based tools. Looking ahead, future human-AI pair programming systems could greatly benefit from seamless integrations and multi-screen environments that move beyond traditional setups, which may enrich collaboration and enhance learning outcomes.

\begin{acks}
We thank our anonymous reviewers for their reviews. This work is supported by the National Science Foundation under award no. NSF-2418582, OAC-2246035, OAC-2441804, as well as OpenAI's Researcher Access Program. 
\end{acks}

\bibliographystyle{ACM-Reference-Format}
\bibliography{main}

\end{document}
\endinput